\documentclass[aps,prd,nofootinbib,amsmath,amssymb,superscriptaddress,showpacs,showkeys,twocolumn,10pt]{revtex4}

\pdfoutput=1

\usepackage{graphicx}
\usepackage{dcolumn}
\usepackage{bm}
\usepackage{amssymb}
\usepackage{latexsym}
\usepackage{booktabs}
\usepackage{amsmath}
\usepackage{multirow}
\usepackage[colorlinks=true, linkcolor=red, citecolor=blue]{hyperref}
\usepackage{color}
\usepackage[usenames,dvipsnames]{xcolor}

\definecolor{darkgreen}{cmyk}{0.85,0.2,1.00,0.2}
\definecolor{purple}{cmyk}{0.5,1.0,0,0}

\newcommand{\LCDM}{$\Lambda$CDM}
\newcommand{\ILCDM}{I$\Lambda$CDM}

\def\msum{$\sum m_\nu$}

\def\({\left(}
\def\){\right)}

\begin{document}

\title{Dark energy versus modified gravity: Impacts on measuring neutrino mass}

\author{Ming-Ming Zhao}
\affiliation{Department of Physics, College of Sciences,
Northeastern University, Shenyang 110819, China}
\author{Rui-Yun Guo}
\affiliation{Department of Physics, College of Sciences,
Northeastern University, Shenyang 110819, China}
\author{Dong-Ze He}
\affiliation{Department of Physics, College of Sciences,
Northeastern University, Shenyang 110819, China}
\author{Jing-Fei Zhang}
\affiliation{Department of Physics, College of Sciences,
Northeastern University, Shenyang 110819, China}
\author{Xin Zhang\footnote{Corresponding author}}
\email{zhangxin@mail.neu.edu.cn}
\affiliation{Department of Physics, College of Sciences,
Northeastern University, Shenyang 110819, China}
\affiliation{Ministry of Education's Key Laboratory of Data Analytics and Optimization
for Smart Industry, Northeastern University, Shenyang 110819, China}
\affiliation{Center for High Energy Physics, Peking University, Beijing 100080, China}

\begin{abstract}
In this paper, we make a comparison for the impacts of smooth dynamical dark energy, modified gravity, and interacting dark energy on the cosmological constraints on the total mass of active neutrinos. For definiteness, we consider the \LCDM~model, the $w$CDM model, the $f(R)$ model, and two typical interacting vacuum energy models, i.e., the \ILCDM1 model with $Q=\beta H\rho_{\rm c}$ and the \ILCDM2 model with $Q=\beta H\rho_{\Lambda}$. In the cosmological fits, we use the Planck 2015 temperature and polarization data, in combination with other low-redshift observations including the baryon acoustic oscillations, the type Ia supernovae, the Hubble constant measurement, and the large-scale structure observations, such as the weak lensing as well as the redshift-space distortions. Besides, the Planck lensing measurement is also employed in this work. We find that, the $w$CDM model favors a higher upper limit on the neutrino mass compared to the $\Lambda$CDM model, while the upper limit in the $f(R)$ model is similar with that of the $\Lambda$CDM model.
For the interacting vacuum energy models, the \ILCDM1 model favors a higher upper limit on neutrino mass, while the \ILCDM2 model favors an identical neutrino mass with the case of $\Lambda$CDM.


\end{abstract}


\maketitle
\section{Introduction}
Neutrino oscillation phenomena {has indicated} that the neutrinos have {nonzero} masses (see Ref.~\cite{Lesgourgues:2006nd} for a review). {However, it is} a great challenge for laboratory experiments of particle physics to measure the absolute neutrino mass scale. The neutrino oscillation experiments can only provide the information about the squared mass differences between the neutrino mass eigenstates \cite{PDG2015-neutrino}, which leads to a lower limit on the species-summed neutrino mass of $\sim 0.06$ eV. Though the tritium beta decay experiments are potentially capable of directly measuring the neutrino mass, the current best upper bound is still weak: $m_\beta<2.3$ eV at 95\% confidence level (C.L.) \cite{beta2,beta1}, where $m_\beta$ is the mass to which the beta decay experiments are sensitive. {The future beta decay experiment, Karlsruhe Tritium Neutrino Experiment} (KATRIN), aims to measure $m_\beta$ with a sensitivity of 0.2 eV \cite{Wolf:2008hf}, which would provide the limit {on the total neutrino mass} $\sum m_\nu\lesssim 0.6$ eV. Additionally, the future experiments for detecting cosmic relic neutrinos, such as the PTOLEMY proposal~\cite{Betts:2013uya,Zhang:2015wua,Huang:2016qmh,Zhang:2017ljh}, are also able to measure the absolute masses of neutrinos. {Currently, the strong constraints on neutrino masses are actually offered by the cosmological observations} (see, e.g., Refs.~\cite{Ade:2013zuv,Ade:2015xua} and references therein).

Actually, the cosmological constraints on the total neutrino mass $\sum m_\nu$ (summed over the three active neutrino species) largely depend on the choices of cosmological models and observational datasets. Massive neutrinos can affect the cosmological expansion rate, and also can lead to a large free-streaming scale (due to their large thermal velocity) below which the clustering amplitude is suppressed. Thus, the observations of cosmic microwave background (CMB) and large-scale structure (LSS) could be used to constrain the neutrino mass. If the total neutrino mass would exceed the limit $\sum m_\nu \lesssim 1.8$ eV, then the CMB observations alone could tightly constrain $\sum m_\nu$ \cite{Dodelson:1995es,Ichikawa:2004zi}. In order to go beyond this limit, one needs to consider more subtle effects of massive neutrinos on the CMB anisotropies, and needs to combine the CMB data with other low-redshift cosmological probes.

The {\it Planck} satellite mission has provided extremely accurate measurements of the CMB anisotropies, with which the spatially flat $\Lambda$CDM cosmology is shown to be highly consistent \cite{Ade:2013zuv,Ade:2015xua}. Based on the standard $\Lambda$CDM model, the Planck 2015 temperature and polarization power spectra in combination with the low-$\ell$ temperature-polarization data (denoted as ``lowP''), gave a better constraint on the species-summed mass of neutrinos: $\sum m_\nu<0.49$ eV (Planck TT,TE,EE + lowP) \cite{Ade:2015xua}. When the baryon acoustic oscillation (BAO) data were added, the limit became much tighter: $\sum m_\nu<0.17$ eV (Planck TT,TE,EE + lowP + BAO) \cite{Ade:2015xua}. When further considering the redshift space distortion (RSD) measurement of LSS from the DR12 galaxy samples in the data combination, the constraint of the neutrino mass was still consistent with the previous results: $\sum m_\nu<0.16$ eV (Planck TT,TE,EE + lowP + BAO + RSD) \cite{Alam:2016hwk}. However, if replacing the RSD data with the galaxy weak lensing from Dark Energy Survey Year 1 (DESY1) combined with the type Ia supernova (SN) measurement, the constraint of neutrino mass would be pulled towards a higher value, which was confirmed by the result in Ref.~\cite{Abbott:2017wau}: $\sum m_\nu<0.29$ eV (Planck TT+BAO+SN+DESY1). Note here that all the upper limit values for neutrino mass quoted in this paper refer to the $95\%$ C.L.

In addition to the dependence of cosmological observational datasets, the measurement of neutrino mass in cosmology also significantly depends on the {assumptions} of cosmological models. Recently, a number of investigations about weighing neutrinos in different dark energy models using cosmological observations have been performed. {In light of these studies \cite{Zhao:2016ecj,Vagnozzi:2018jhn,Zhang:2015uhk,Chen:2015oga,Zhang:2015rha,Zhao:2017urm,Feng:2017mfs}, some essential conclusions are drawn and summarized} in Ref.~\cite{Zhang:2017rbg}: (1) In the dynamical dark energy models, {compared to the $\Lambda$CDM case, the upper limit of $\sum m_\nu$ could become either larger or smaller}, depending on the nature of dark energy. It is found that, in the cases of a phantom dark energy or an early phantom dark energy (i.e., $w<-1$ in the early times), the constraint on $\sum m_\nu$ becomes looser, but in the cases of a quintessence dark energy or an early quintessence dark energy (i.e., $-1<w<-{1}/{3}$ in the early times), the constraint on $\sum m_\nu$ becomes tighter. (2) In the holographic dark energy (HDE) model \cite{Li:2004rb,hde2,Zhang:2006qu,Zhang:2005hs,Chang:2005ph,Zhang:2006av,Zhang:2007sh,Zhang:2007an,Zhang:2007es,Ma:2007av,Zhang:2014ija}, one gets the tightest constraint on the neutrino mass, namely $\sum m_\nu<0.105$ eV, which is almost equal to the lower bound of $\sum m_\nu$ for the inverted hierarchy case. (3) The mass splitting of neutrinos {is able to} influence the cosmological fits. Moreover, the normal hierarchy case can fit the current observations slightly better than the inverted hierarchy case, which is independent of dark energy models. Of course, these statements still need to be further carefully checked in the upcoming future.

In fact, while the acceleration of the cosmic expansion was discovered by the {observations of type Ia supernovae} and was {further} confirmed by {a series of} other independent astronomical/astrophysical observations, the physics behind this {counterintuitive phenomenon} is still obscure (see Ref.~\cite{Mortonson:2013zfa} for a brief review). If the general relativity (GR) is valid on all the scales in the universe, one needs to introduce a new form of {perfect fluid} whose pressure is negative (yielding a repulsive gravity), referred to as ``dark energy'' (DE), to account for the origin of the cosmic acceleration. {Particularly, the cosmological constant $\Lambda$ (vacuum energy) with $w=-1$ (here, $w$ is the equation-of-state parameter defined by $w=p/\rho$) is the physically simplest candidate for DE. Although favored by the observations, it always suffers from the fine-tuning and coincidence problems. To solve or alleviate these theoretical puzzles, many other possible theoretical models have been proposed. For example, the spatially homogeneous, slowly rolling scalar field, dubbed ``quintessence'', provides a possible mechanism for dynamical dark energy}. Nevertheless, there also exists another possibility that GR would be violated on the cosmological scales, which leads to a special approach to explain the accelerated cosmic expansion through modifying the gravity (see, e.g., Refs.~\cite{Sotiriou:2008rp,DeFelice:2010aj,Clifton:2011jh,Weinberg:2012es,Joyce:2014kja} for recent reviews). In other words, the modified gravity (MG) theory could mimic the  {``effective dark energy'' at the cosmological background level to drive the cosmic acceleration}. One typical example for MG is to replace the Ricci scalar $R$ with a function $R+f(R)$ in the gravitational action \cite{Carroll:2003wy}. {Moreover, a more radical idea to construct MG is to introduce the extra dimensions in braneworld theories, allowing gravitons to ``leak'' off the brane that represents our observable universe} (e.g., the DGP model \cite{DGP}), which has inspired a more general class of ``galileon'' and massive gravity models.

{Within the framework of GR}, the histories of cosmic structure growth and cosmic expansion are linked by a consistency relation, $\ddot{D}+2H\dot{D}-4\pi G \rho_{\rm m} D=0$, where $D(z)$ is the linear growth function, $H(z)$ is the Hubble expansion rate, and $G$ is the Newton's gravitational constant. MG models can change the predicted linear growth function, and typically make the growth function dependent on the scale or environment, i.e., $D(z,k)$. Thus, {the violation of consistency relation for GR would occur in the MG models, leading to the consequence that the growth index $\gamma$ deviates from its GR value}, $\gamma\approx 0.55$ \cite{Wang:1998gt,Linder:2005in}. In addition, the MG models can typically yield a non-vanishing anisotropic stress (proportional to $\Phi-\Psi$, where $\Phi$ and $\Psi$ are two gauge invariant potentials describing the scalar metric perturbations).

{Now that the determination of neutrino mass depends on the DE models in cosmology, it is of great interest to quantify the impacts of the DE models explaining the cosmic acceleration with various theoretical motivations (dark energy and modified gravity) on the constraint results of \msum. As there are too many seemingly reasonable models of DE and MG resulting from multiple physical considerations, it is of course not possible and not necessary to exhaust all the models in this work. For simplicity, we only choose the most typical (or simplest) models as the examples to do the analysis. For the smooth dynamical dark energy, we consider the $w$CDM model that modifies the background equations of cosmological constant but yields a similar structure growth history to the $\Lambda$CDM model \cite{Linder:2005in}. For the modified gravity, we consider a parameterized $f(R)$ model whose background is fixed as that of the $\Lambda$CDM model, with the perturbations parameterized by $B_0$ that characterizes the Compton wavelength scale for the extra scalar degree of freedom} \cite{Song:2006ej} (see also Refs.~\cite{Bertschinger:2008zb,Giannantonio:2009gi,Li:2015fR}).

{However, it must be pointed out that there is another class of DE model, in which some subtle direct interaction exists between dark energy and cold dark matter, called interacting dark energy (IDE) model} \cite{Amendola:1999er,Zimdahl:2005bk,Wang:2006qw,Guo:2007zk,Koyama:2009gd,Wei:2010cs,He:2010im,Li:2011ga,Zhang:2012uu,Zhang:2013lea,Li:2013bya,Geng:2015ara,Murgia:2016ccp,Wang:2016lxa,Pourtsidou:2016ico,Feng:2016djj,Xia:2016vnp,Costa:2016tpb,vandeBruck:2016hpz,Kumar:2017dnp,Sola:2017jbl,Zhang:2005rg,Zhang:2005rj,Zhang:2004gc,Zhang:2007uh,Li:2009zs,Zhang:2009qa,Wang:2014oga,Guo:2017deu,Li:2017usw,Li:2018ydj}. This scenario is different from MG in that the so called ``fifth force'' only exists between dark energy and cold dark matter. Usually, in the scenario of IDE, both of the histories of cosmic expansion and the growth of structure differ from the \LCDM\ universe. {What is important is that the models of IDE can successfully resolve (or alleviate) the fine-tuning and coincidence problems through the attractor solution}. Recently, some works related to the constraints on the total neutrino mass in the IDE models using current cosmological observations have been done~\cite{Guo:2017hea,Guo:2018gyo,Feng:2019mym,Feng:2019jqa}. {In the present work, we as well take the IDE scenario into consideration so as to make the research more general and comprehensive. In order to introduce the fewest parameters}, we only consider the simplest model of IDE, i.e., the interacting vacuum energy model, in which the interaction between vacuum energy and cold dark matter is assumed. {For convenience, we call this kind of IDE model the I$\Lambda$CDM model, although the vacuum energy density is no longer a pure background in this scenario. We consider another two cases of energy transfer rate, $Q=\beta H\rho_{\Lambda}$ and $Q=\beta H\rho_{\rm c}$, with $\rho_{\Lambda}$ the energy density of vacuum energy and $\rho_{c}$ the energy density of cold dark matter}. {In this work, we aim to investigate how the cosmological constraints on the sum of the neutrino masses are affected by the assumptions of the underlying scenarios accounting for the cosmic accelerated expansion, i.e., the $\Lambda$CDM, $w$CDM, $f(R)$, and I$\Lambda$CDM models.}

The rest of this paper is organized as follows: In Sec.~\ref{model}, we will briefly describe the cosmological models considered in this work. In Sec.~\ref{data}, we will introduce the analysis method and observational data employed in this work. In Sec.~\ref{sec:results}, we will report the constraint results in the various cosmological models and make some relevant discussions.  The conclusion will be given in Sec.~\ref{sec:discussion}.

\section{Model}\label{model}

In this section, we shall briefly describe the cosmological models considered in this work, namely the $w$CDM model, the $f(R)$ model, and the I$\Lambda$CDM model.
Here we only provide the basic information of them.

\subsection{The $w$CDM model}

The $w$CDM model is the simplest extension to the $\Lambda$CDM model, and thus is the {naivest candidate for a dynamical dark energy}, in which {one assumes that the equation-of-state parameter of DE is $w=\rm constant$}. Although it is hard to believe that such a simple model with a constant $w$ would correspond to {the final physical situation of} dark energy, it is still very useful in exploring the nature of dark energy with the cosmological parameter estimation due to its simplicity. In this model, we have
\begin{equation}\label{equation}
  E^2(z)=\Omega_{\rm m}(1+z)^{3}+(1-\Omega_{\rm m})(1+z)^{3(1+w)},
\end{equation}
where $E(z)=H(z)/H_{0}$ and $\Omega_{\rm m}$ is the present-day fractional energy density of matter. The radiation density in the late universe is negligible, and thus we do not show it here.

\subsection{The $f(R)$ model}

The $f(R)$ gravity is a simple and nontrivial extension of GR, which has received much attention as an alternative mechanism of $\Lambda$CDM.
In general, the model of $f(R)$ is derived by adding a function of the Ricci scalar $R$ to the Einstein-Hilbert action \cite{Carroll:2003wy}.

By this, an extra scalar degree of freedom $f_{R}\equiv df/dR$ is introduced in the cosmological model. We characterize this function using a dimensionless quantity $B$ \cite{Song:2006ej}, which can be written as
\begin{equation}\label{quantity}
  B\equiv \frac{f_{RR}}{1+f_{R}}\frac{dR}{d\ln a}(\frac{d \ln H}{d \ln a})^{-1},
\end{equation}
where $H$ is the Hubble expansion rate, $a$ is the scale factor, and $f_{RR}$ is the derivative to $f_{R}$ with respect to $R$, i.e., $f_{RR}\equiv d^{2}f/d R^{2}$.

Here, we only consider a parameterized $f(R)$ model whose background is fixed as that of $\Lambda$CDM and perturbations are parameterized by $B_0$ that characterizes the Compton wavelength scale for the extra scalar degree of freedom (see Refs.~\cite{Bertschinger:2008zb,Giannantonio:2009gi,Li:2015fR}),
\begin{equation}\label{mu}
\mu(k,a)=\frac{1}{1-1.4\cdot10^{-8}|\lambda_{1}|^{2}a^{3}}\frac{1+4/3\lambda_{1}^{2}k^{2}a^{4}}{1+\lambda_{1}^{2}k^{2}a^{4}},
\end{equation}
\begin{equation}\label{gama}
  \gamma(k,a)=\frac{1+2/3\lambda_{1}^{2}k^{2}a^{4}}{1+4/3\lambda_{1}^{2}k^{2}a^{4}},
\end{equation}
where the parameter $\lambda_{1}$ has dimension of length, relating to the $B$ value today through $\lambda_{1}^{2}=B_{0}c^{2}/2H_{0}^{2}$. The usages of the functions $\mu(k,a)$ and $\gamma(k,a)$ in the public {\tt MGcamb} code and the details of how to quantify the modifications to the Poisson and anisotropy equations can be found in Refs.~\cite{Lewis:1999bs,Hojjati:2011ix}.

\subsection{The I$\Lambda$CDM model}

{The I$\Lambda$CDM model is the simplest interacting dark energy model, in which the vacuum energy directly interacts with cold dark matter (i.e., the vacuum energy could decay into cold dark matter or vice versa).}

In this model, the energy continuity equation for the vacuum energy density is given by
\begin{equation}\label{vacuum}
  \dot{\rho}_{\Lambda}=Q,
\end{equation}
where the dot represents the derivative with respect to the cosmic time $t$, and $Q$ is the energy (density) transfer rate. For the cold dark matter density, the energy continuity equation reads,
\begin{equation}\label{coldmatter}
  \dot{\rho}_{\rm c}=-3H\rho_{\rm c}-Q.
\end{equation}

It should be mentioned that the energy transfer rate $Q$ can only be given by purely phenomenological consideration. {This is because actually we know little about how dark matter perceives the ``fifth force" through the mediation of some unknown scalar field degrees of freedom of dark energy in a realistic physical mechanism}. In this work, we choose two simple phenomenological forms of energy transfer rate, i.e., {$Q=\beta H\rho_{\rm c}$ (for \ILCDM1\ model) and $Q=\beta H\rho_{\Lambda}$ (for \ILCDM2\ model)}, where $\beta$ denotes a dimensionless coupling parameter. From Eqs.~(\ref{vacuum}) and (\ref{coldmatter}), one can easily see that, $\beta>0$ indicates that the energy transfer is from cold dark matter to vacuum energy; {$\beta<0$ represents an inverse energy transfer orientation}; $\beta=0$ indicates that there is no interaction between cold dark matter and vacuum energy, and {the model recovers to the standard $\Lambda$CDM universe}.

In a perturbed universe, there is also momentum transfer between vacuum energy and cold dark matter, which leads to the large-scale instability problem \cite{Valiviita:2008iv} in the I$\Lambda$CDM cosmology. To solve this problem, we adopt the extended version of parameterized post-Freidmann (PPF) approach in which the interacting dark energy models are accommodated. For details about this version of PPF, we refer the reader to Refs.~\cite{Li:2014eha,Li:2014cee,Li:2015vla,Zhang:2017ize,Guo:2017hea,Feng:2017usu,Guo:2018gyo,Feng:2018yew}.

In this work, {we treat the $\Lambda$CDM model as a reference model. The $w$CDM model, as a representative of smooth dark energy, has the similar history of the growth of structure but different expansion history; the $f(R)$ model, as a representative of MG, has the similar expansion history but different structure growth history; the I$\Lambda$CDM model, as a representative of interacting dark energy, has different evolutionary history of not only cosmic expansion but also structure growth. All of these models have only one additional parameter compared with the $\Lambda$CDM model, i.e., $w$ for the $w$CDM model, $B_0$ (that is replaced by its logarithmic form of $\log_{10}B_{0}$ in calculation, following Ref.~\cite{Dossett:2014oia}) for the $f(R)$ model, and $\beta$ for the two typical I$\Lambda$CDM models. Therefore, it is meaningful and equitable to compare the results of weighing neutrinos in the $\Lambda$CDM, $w$CDM, $f(R)$, and I$\Lambda$CDM models.}

\section{Method and data}\label{data}

The basic cosmological parameters for the 6-parameter {base} $\Lambda$CDM model are $\{{\Omega_{\rm b}h^{2},\Omega_{\rm c}h^{2},\theta_\ast,\tau,A_{\rm s},n_{\rm s}}\}$. These parameters, one by one, represent the energy density of baryons, the energy density of cold dark matter, {the acoustic angular size of the sound horizon at the time of last scattering}, the reionization optical depth, {the power amplitude of the primordial curvature perturbations}, and the tilt of the primordial scalar fluctuations, {respectively}. When the massive neutrinos are considered, the neutrino mass parameter $\sum m_{\nu}$ should also be included in all {the considered} models. Recently, the issue concerning the mass hierarchy of neutrinos has been discussed in  Refs.~\cite{Huang:2015wrx,Wang:2016tsz,Guo:2018gyo,Zhao:2017jma}. But in the present work we do not consider the mass splittings of neutrinos, and thus three mass-degenerate neutrino species are assumed.

We employ several significant cosmological observational data sets to perform our analysis, which include the CMB temperature and polarization power spectra, BAO, SN, $H_{0}$, redshift-space distortion (RSD), and weak lensing (WL) data.

\begin{itemize}
 \item The CMB data: We use the combination of the Planck 2015 temperature and polarization power spectra at multipoles $\ell>30$ and low-$\ell$ temperature+polarization likelihood \cite{Aghanim:2015xee}.
 \item The BAO data: We use the combination of the previous measurements from 6dFGS ($z=0.1$) \cite{Beutler:2011hx}, SDSS-MGS ($z=0.15$) \cite{Ross:2014qpa}, and the latest measurements from LOWZ ($z=0.32$) and CMASS ($z=0.57$) DR12 samples of BOSS \cite{Gil-Marin:2015nqa}.
 \item The SN data: For the type Ia supernova observation, we adopt the ``JLA" sample, compiled from the SNLS, SDSS and the samples of several low-redshift SN data~\cite{Betoule:2014frx}.
 \item The $H_{0}$ data: We use the direct measurement of $H_0$ reanalysed by Efstathiou \cite{Efstathiou:2013via}, which is rather consistent with the recent results of $H_{0}$ from the Planck 2015 data \cite{Ade:2015xua}. The consistent measurement value of the Hubble constant is $H_{0}=70.6\pm3.3 ~\rm km  ~s^{-1} ~Mpc^{-1}$.
 \item The RSD data: We employ two RSD measurements from BOSS CMASS DR12 ($z=0.57$) and LOWZ DR12 ($z=0.32$) samples \cite{Gil-Marin:2016wya}, respectively. The usage of the RSD data is the same as the prescription given by the Planck collaboration \cite{Ade:2015xua}. {Because these two RSD data  include the corresponding BAO information, we do not consider the BAO measurements \cite{Gil-Marin:2015nqa} from the BAO likelihood in the context of employment of the RSD data to avoid double counting}.

 \item The WL data: We use a measurement of $S_{8}=\sigma_{8}(\Omega_{\rm m}/0.3)^{0.5}$, derived by the weak gravitational lensing from the Dark Energy Survey Year 1 (DESY1) survey, with $S_{8}=\sigma_{8}(\Omega_{\rm m}/0.3)^{0.5}=0.783^{+0.021}_{-0.025}$ ($2\sigma$) \cite{Abbott:2017wau}. Since this measurement result slightly deviates from a gaussian form, we simply make it gaussian by taking the average of the errors, {(0.021+0.025)/2}, i.e., $S_{8}=0.783\pm0.023$.
 \item The lensing data: We also use the CMB lensing power spectrum from the Planck lensing measurement \cite{Ade:2015zua}. We denote the CMB lensing measurement as ``lensing''.

\end{itemize}

We employ the {\tt CosmoMC} package \cite{Lewis:2002ah} to infer the posterior probability distributions of parameters. {We utilize flat priors for the basic cosmological parameters, and the prior ranges for the parameters are set to be much wider than the posterior in order not to affect the results of parameter estimation.}  We also employ the {\tt MGcamb} code \cite{Lewis:1999bs,Hojjati:2011ix} {in particular} to do the calculations for the $f(R)$ model.

\section{Results}\label{sec:results}

\subsection{Neutrino mass}

\begin{figure}
\begin{center}
\includegraphics[scale=0.8, angle=0]{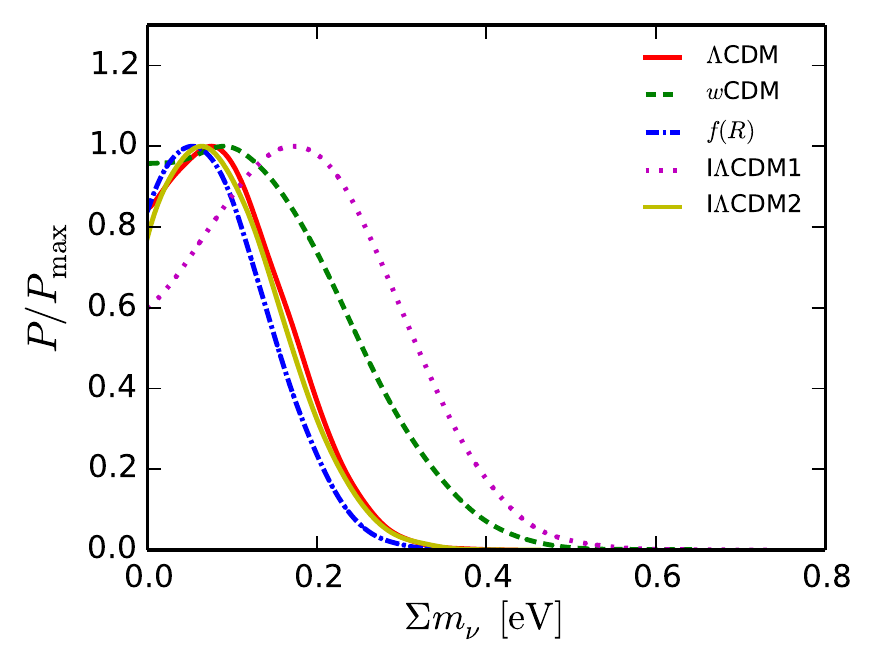}
\caption{One-dimensional joint, marginalized distributions of $\sum m_{\nu}$ for the $\Lambda$CDM model (red solid), the $w$CDM model (green dashed), the $f(R)$ model (blue dashed-dotted), the I$\Lambda$CDM1 ($Q=\beta H\rho_{\rm c}$) model (purple dotted), and the I$\Lambda$CDM2 ($Q=\beta H\rho_{\Lambda}$) model (yellow solid) from the constraints of the CMB+BAO+SN+$H_{0}$+WL+RSD+lensing data combination.}
\label{fig:mnu}
\end{center}
\end{figure}

We constrain the total mass of neutrinos in the models of $w$CDM, $f(R)$, I$\Lambda$CDM1 ($Q=\beta H\rho_{\rm c}$), and I$\Lambda$CDM2 ($Q=\beta H\rho_{\Lambda}$) by using the data combination of CMB+BAO+SN+$H_{0}$+WL+RSD+lensing. {Note here that for the employment of ``$H_{0}$ prior", we do not adopt the local determination of the Hubble constant ($H_{0}=73.00\pm1.75 ~\rm km  ~s^{-1} ~Mpc^{-1}$) by Reiss {et al}. \cite{Riess:2016jrr} for two causes: (1) There exists beyond $3\sigma$ tension with Planck measurement, and (2) a higher $H_{0}$ prior is more prone to derive a smaller $\sum m_{\nu}$ due to the strong anti-correlation between $H_{0}$ and $\sum m_{\nu}$. Hence, we choose a relatively lower prior of $H_{0}$ re-analyzed by Efstathiou \cite{Efstathiou:2013via}, $H_{0}=70.6\pm3.3 ~\rm km  ~s^{-1} ~Mpc^{-1}$, which is well consistent with Planck measurement. To visually show the impacts of these cosmological models on the constraints on the neutrino mass, we plot the one-dimensional joint, marginalized distributions of $\sum m_{\nu}$ for the $\Lambda$CDM model (red solid line), the $w$CDM model (green dashed line), the $f(R)$ model (blue dashed-dotted line), the I$\Lambda$CDM1 model (purple dotted line), and the I$\Lambda$CDM2 model (yellow solid line) one by one, as shown in Fig.~\ref{fig:mnu}. We also exhibit the detailed results in Table~\ref{tab1}.}

{In this work we take the $\Lambda$CDM model as a reference model. In the $\Lambda$CDM model, the 95\% C.L. upper limit on the sum of the neutrino masses, $\sum m_{\nu}<0.22$ eV, is obtained.
Importantly, we are supposed to pay special attention to the constraints on the neutrino mass in other DE and MG models. As shown in Table~\ref{tab1}, for the $w$CDM model, the upper limit $\sum m_{\nu}<0.33$ eV is obtained. For the I$\Lambda$CDM1 model and the I$\Lambda$CDM2 model, the upper limits of $\sum m_{\nu}<0.37$ eV and $\sum m_{\nu}<0.22$ eV, are obtained. For the $f(R)$ model, the upper limit $\sum m_{\nu}<0.20$ eV is obtained. For these results, as vividly displayed in Fig.~\ref{fig:mnu}, we can obviously find that the $w$CDM model and the I$\Lambda$CDM1 model could pull the upper limit on neutrino mass away from the case of the $\Lambda$CDM model, towards a higher value. Nevertheless, the I$\Lambda$CDM2 model and $f(R)$ model indicate the approximately equal upper limits on the neutrino mass to the case of the $\Lambda$CDM model. Thus, according to the fitting results, we can conclude that some models considered in this work can impose a significant influence on weighing neutrinos, such as the $w$CDM model and the I$\Lambda$CDM1 ($Q=\beta H\rho_{\rm c}$) model, but the other models can only exert a marginal impact on weighing neutrinos, such as the I$\Lambda$CDM2 ($Q=\beta H\rho_{\Lambda}$) model and the $f(R)$ model.}

\begin{figure*}
\begin{center}
\includegraphics[scale=0.6, angle=0]{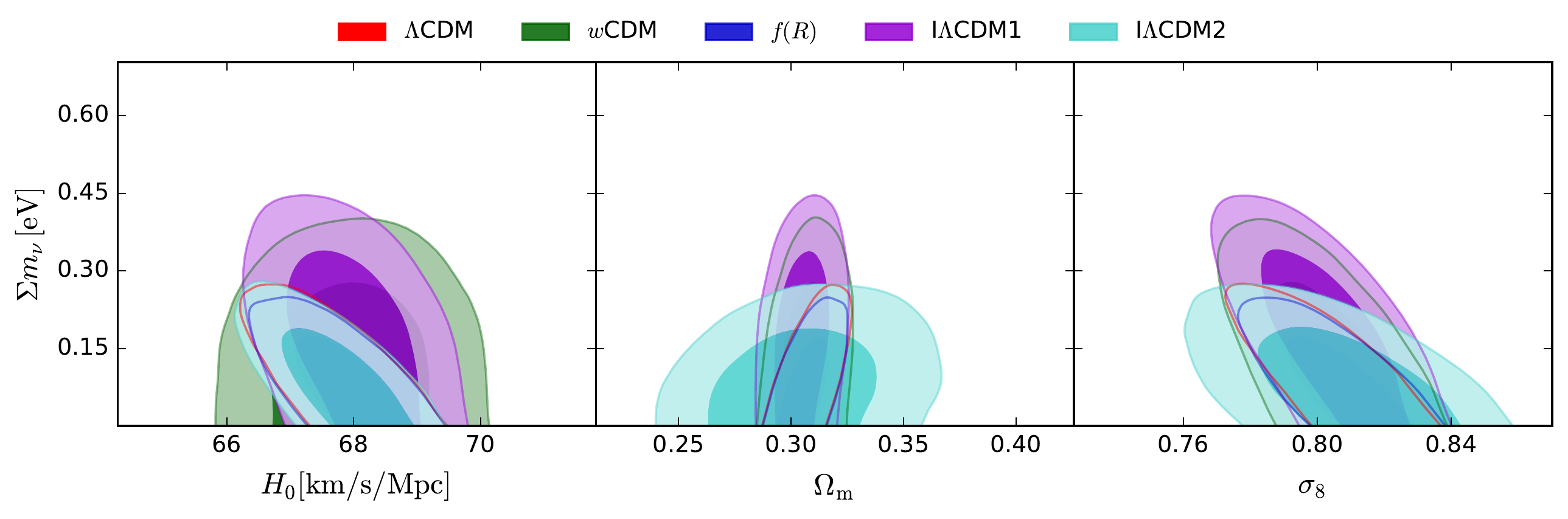}
\caption{The 68\% and 95\% C.L. contours in the parameter planes of $\sum m_{\nu}$ versus $H_{0}$, $\Omega_{\rm m}$ and $\sigma_{8}$ for the $\Lambda$CDM model, the $w$CDM model, the $f(R)$ model, the I$\Lambda$CDM1 ($Q=\beta H\rho_{\rm c}$) model, and the I$\Lambda$CDM2 ($Q=\beta H\rho_{\Lambda}$) model from the constraints of the CMB+BAO+SN+$H_{0}$+WL+RSD+lensing data combination.}
\label{fig:mnu-om-sig-h0}
\end{center}
\end{figure*}

In addition, we also compare the constraint results of other late-time cosmological parameters strongly correlated with the neutrino mass in these cosmologies. Fig.~\ref{fig:mnu-om-sig-h0} shows the 68\% C.L. and 95\% C.L. constraints on neutrino mass $\sum m_{\nu}$ plus the late-time parameters $H_{0}$, $\Omega_{\rm m}$, and $\sigma_{8}$ for the considered five models respectively. {We find that the best fit values of $H_{0}$, $\Omega_{\rm m}$, and $\sigma_{8}$, as well as the derived correlations between the neutrino mass and the late-time parameters are roughly consistent in these five models.}
However, the parameter space distributions of the late-time parameters are greatly different. It is found that compared with the $\Lambda$CDM model, the $w$CDM model and the I$\Lambda$CDM1 model yield rather broad distributions of $H_{0}$ in the $H_{0}-\sum m_{\nu}$ plane. Besides, in the $\Omega_{\rm m}-\sum m_{\nu}$ plane and the $\sigma_{8}-\sum m_{\nu}$ plane, the I$\Lambda$CDM2 model can yield broader distributions of $\Omega_{\rm m}$ and $\sigma_{8}$ compared with the other models. Moreover, it is of interest to notice that the constraints of these parameters for $f(R)$ are highly consistent with the case of $\Lambda$CDM.

\begin{table*}
\caption{\label{tab1} Fitting results for the $\Lambda$CDM, $w$CDM, $f(R)$, I$\Lambda$CDM1 ($Q=\beta H\rho_{\rm c}$) and I$\Lambda$CDM2 ($Q=\beta H\rho_{\Lambda}$) {models} with the CMB+BAO+SN+$H_{0}$+WL+RSD+lensing data combination. Here, we fix $N_{\rm eff}=3.046$. We quote the $\pm 1\sigma$ errors, but for the neutrino mass $\sum m_{\nu}$ and $\log_{10}B_{0}$, we quote the 95\% C.L. upper limits.}
\centering
\renewcommand{\arraystretch}{1.3}
\begin{tabular}{cccccccccc}\hline\hline
 Model&$\Lambda$CDM&$w$CDM&$f(R)$&I$\Lambda$CDM1 &I$\Lambda$CDM2 \\
\hline
$\sum m_\nu$ [eV]&$<0.22$&$<0.33$&$<0.20$&$<0.37$&$<0.22$\\
$w$/$\log_{10}B_{0}$/$\beta$&$-$&$-1.042^{+0.071}_{-0.024}$&$<-5.67$&$0.0018\pm0.0002$&$0.0175^{+0.1231}_{-0.1133}$\\
$\Omega_{\rm m}$&$0.3082^{+0.0072}_{-0.0079}$&$0.3076^{+0.0083}_{-0.0082}$&$0.3075^{+0.0067}_{-0.0077}$&$0.3051\pm0.0085$&$0.3050^{+0.0250}_{-0.0280}$\\
$H_0$ [km s$^{-1}$ Mpc$^{-1}$] &$67.74^{+0.65}_{-0.59}$&$67.96\pm0.86$&$67.81^{+0.63}_{-0.54}$&$67.94\pm0.70$&$67.68^{+0.69}_{-0.60}$\\
$\sigma_8$&$0.804^{+0.013}_{-0.011}$&$0.802^{+0.014}_{-0.013}$&$0.806^{+0.012}_{-0.011}$&$0.803^{+0.015}_{-0.013}$&$0.806^{+0.021}_{-0.019}$\\
\hline\hline
\end{tabular}
\end{table*}

\subsection{Model parameters}\label{cosmologymodel}

\begin{figure*}
\begin{center}
\includegraphics[scale=0.6, angle=0]{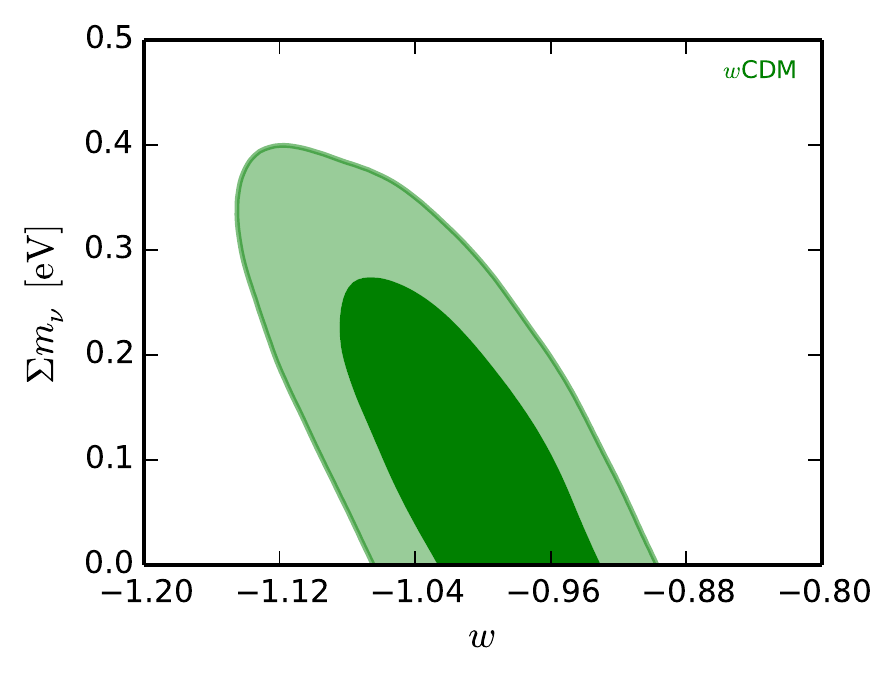}
\includegraphics[scale=0.6, angle=0]{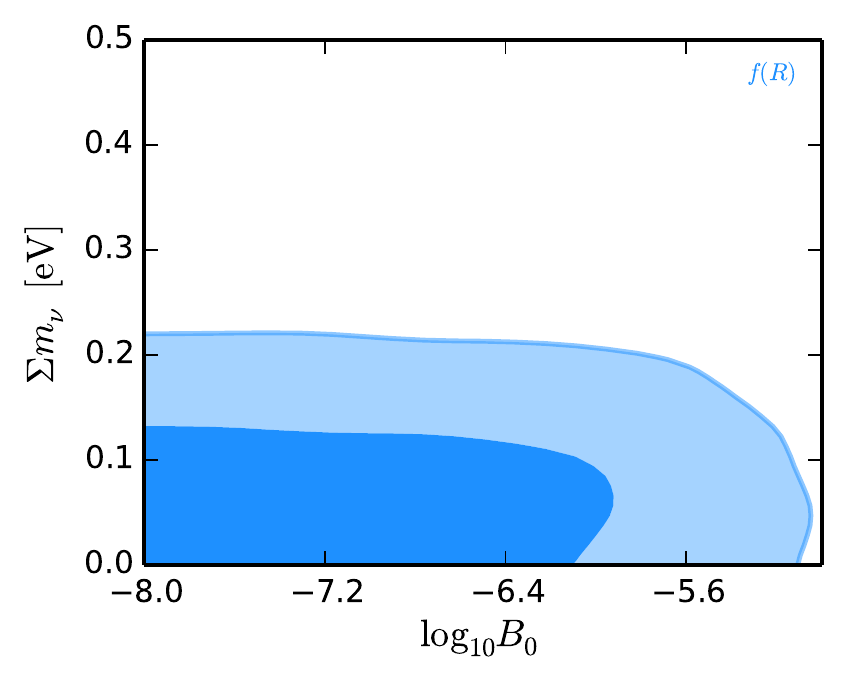}\\
\includegraphics[scale=0.6, angle=0]{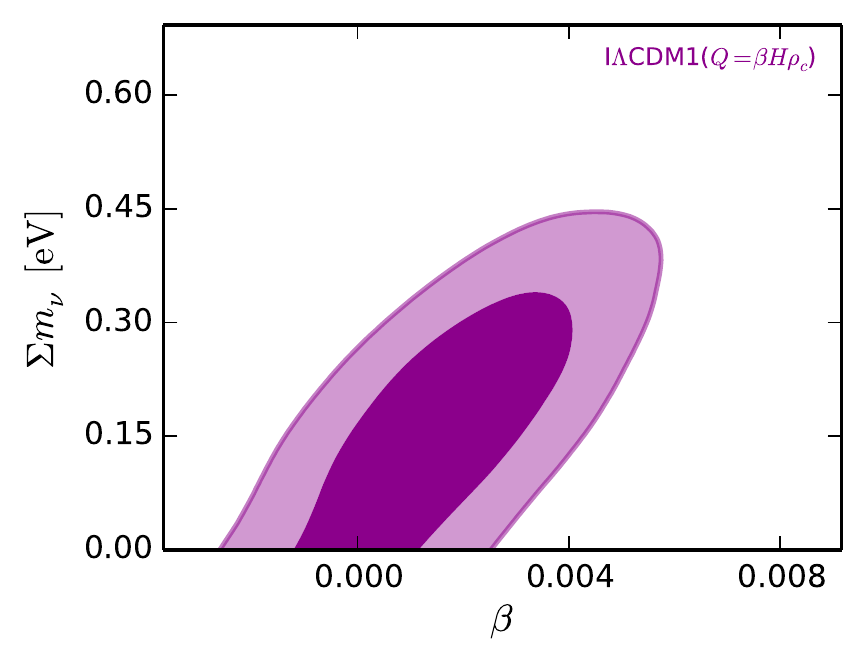}
\includegraphics[scale=0.6, angle=0]{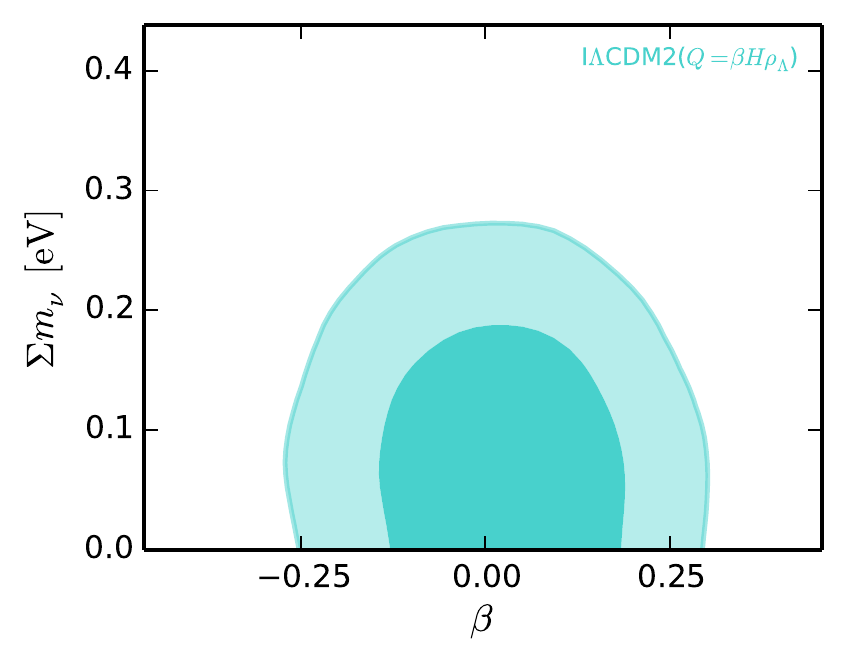}
\caption{The 68\% and 95\% C.L. contours in the $w-\sum m_{\nu}$ plane for the $w$CDM model, the $\log_{10}B_{0}-\sum m_{\nu}$ plane for the $f(R)$ model, the $\beta-\sum m_{\nu}$ plane for the I$\Lambda$CDM1 ($Q=\beta H \rho_{\rm c}$) model, and the $\beta-\sum m_{\nu}$ plane for the I$\Lambda$CDM2 ($Q=\beta H\rho_{\Lambda}$) model.}
\label{fig:all-mnu}
\end{center}
\end{figure*}

In the last subsection, we have found that the $w$CDM, $f(R)$, and I$\Lambda$CDM models possess different effects on the cosmological constraints of neutrino mass. In order to investigate how these models affect the constraints of the neutrino mass, in Fig.~\ref{fig:all-mnu} we plot the two-dimensional posterior distribution contours of the specific model parameters and the neutrino mass in the $w-\sum m_{\nu}$ plane for $w$CDM model, in the $\log_{10}B_{0}-\sum m_{\nu}$ plane for $f(R)$ model, and in the $\beta-\sum m_{\nu}$ plane for I$\Lambda$CDM1 and I$\Lambda$CDM2 models.

From the left two panels in Fig.~\ref{fig:all-mnu}, we can see that $w$ is anti-correlated with $\sum m_{\nu}$ for $w$CDM, and $\beta$ is positively correlated with $\sum m_{\nu}$ for I$\Lambda$CDM1. While for the right two panels in Fig.~\ref{fig:all-mnu}, there are no obvious correlations between $\log_{10}B_{0}$ and $\sum m_{\nu}$ for $f(R)$, and between $\beta$ and $\sum m_{\nu}$ for I$\Lambda$CDM2. We find that, when the model parameter is correlated with the neutrino mass, the upper limit of the sum of the neutrino mass in these models will differ from the case of $\Lambda$CDM. On the contrary, if the model parameter is not correlated with the neutrino mass, the derived limits of the neutrino mass will be well consistent with the case of $\Lambda$CDM.

Actually, the apparently strong correlations between the model parameters and the neutrino mass can be explained by the effects on the shape of the CMB power spectra. The effects on the background cosmology can be compensated by changing some other parameters to guarantee the fixed acoustic angular scale $\theta_{*}$. For example, in the models of $w$CDM and I$\Lambda$CDM1, the evolution of the energy density of DE due to the dynamical properties of DE described by the specific model parameters could impact the expansion rate of universe and thus the angular diameter distance $D_{\rm A}$ ($D_{\rm A}\propto 1/\theta_{*}$), which would be compensated by the shifts of neutrino mass.

On the other hand, some model parameters can only exert a very slight effect on the background cosmology and the shape of the CMB power spectra, leading to the weak correlations between the model parameters and the neutrino mass. For example, in the $f(R)$ model, the model parameter $\log_{10}B_{0}$ is a parameterized form of the perturbations under the MG theory, which is mainly related to the growth of structure. The parameter $\log_{10}B_{0}$ has been tightly constrained by the current observations, and thus according to the current observational constraints the $f(R)$ model is like the $\Lambda$CDM model in both background and perturbation aspects. See Refs.~\cite{Hu:2014sea,Bellomo:2016xhl} for more relevant discussions on the same topic.



For the $w$CDM model, we have $w=-1.042^{+0.071}_{-0.024}$ ($1\sigma$), which means that a phantom energy is mildly favored. 
For the I$\Lambda$CDM1 model, we obtain the coupling constant $\beta=0.0018\pm0.0002$ ($1\sigma$), which implies that the current observations prefer the possibility $\beta>0$ (cold dark matter decays into vacuum energy) at the level of $1\sigma$. For the I$\Lambda$CDM2 model, we obtain $\beta=0.0175^{+0.1231}_{-0.1133}$ ($1\sigma$), which indicates that the result of $\beta=0$ (equivalent to the case of $\Lambda$CDM) is favored by current observations at the $1\sigma$ level. In other words, we find no evidence of an obvious deviation from the $\Lambda$CDM cosmology in this model.
For the $f(R)$ model, we obtain an upper limit $\log_{10}B_{0}<-5.67$ ($2\sigma$), showing that the value of $B_0$ is rather small, and thus we also find no evidence of an obvious deviation from the $\Lambda$CDM cosmology in this model. This is why in the I$\Lambda$CDM2 model and the $f(R)$ model the constraints on the neutrino mass are almost the same as that in the $\Lambda$CDM model.



\begin{figure*}
\begin{center}
\includegraphics[scale=0.5, angle=0]{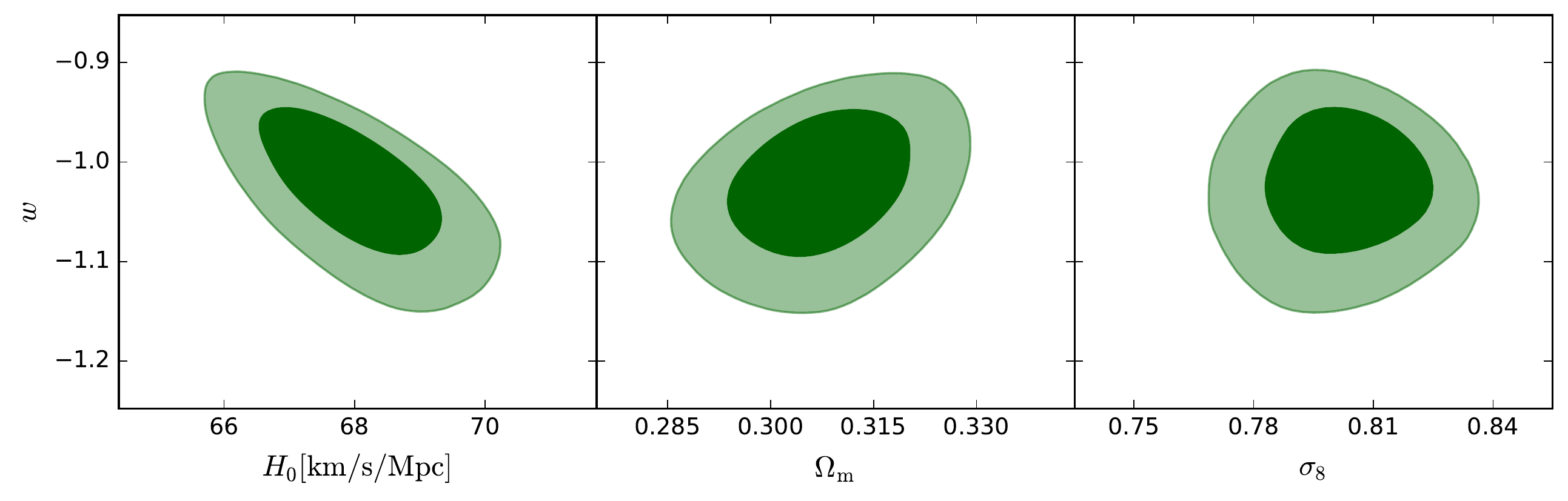}
\includegraphics[scale=0.5, angle=0]{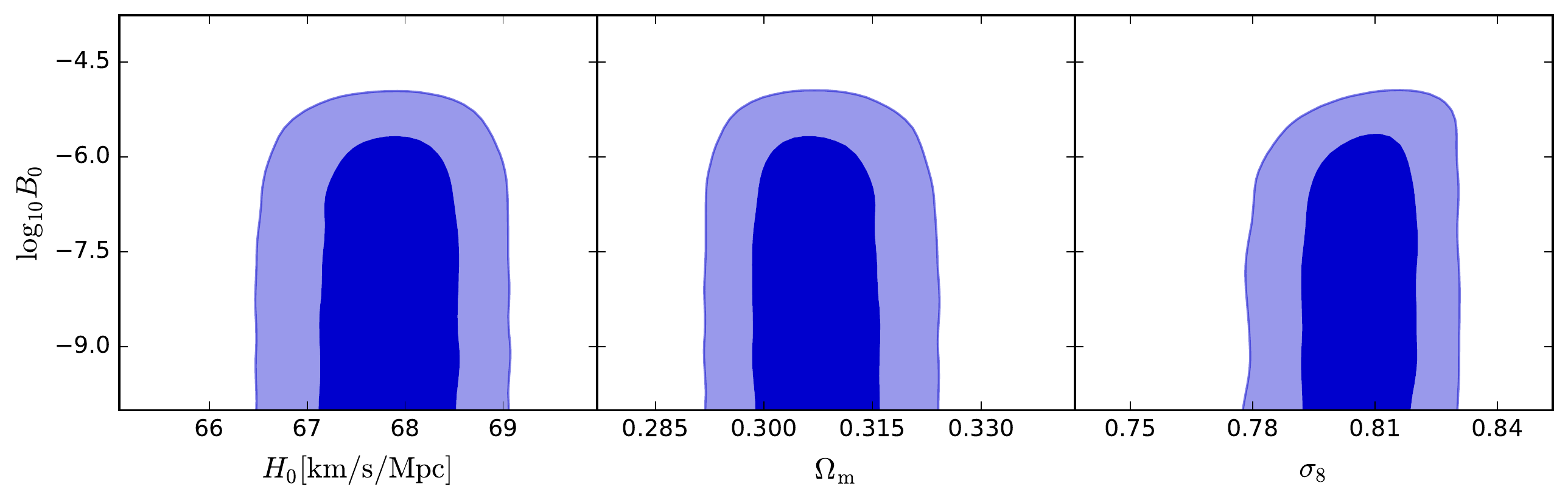}\\
\includegraphics[scale=0.5, angle=0]{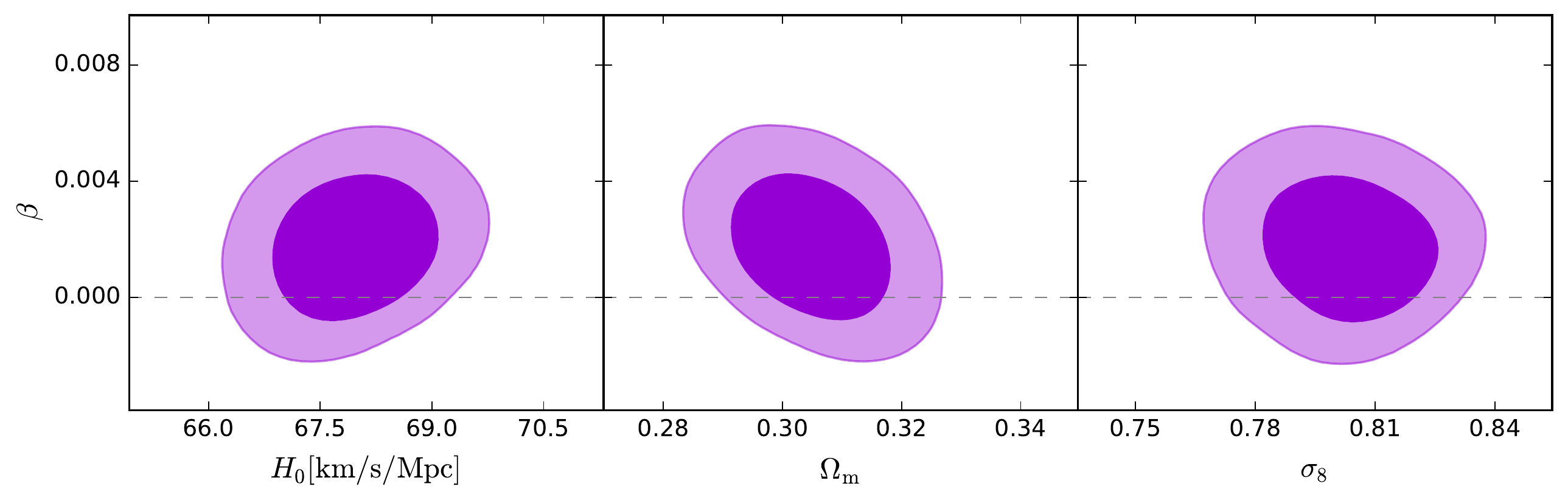}
\includegraphics[scale=0.5, angle=0]{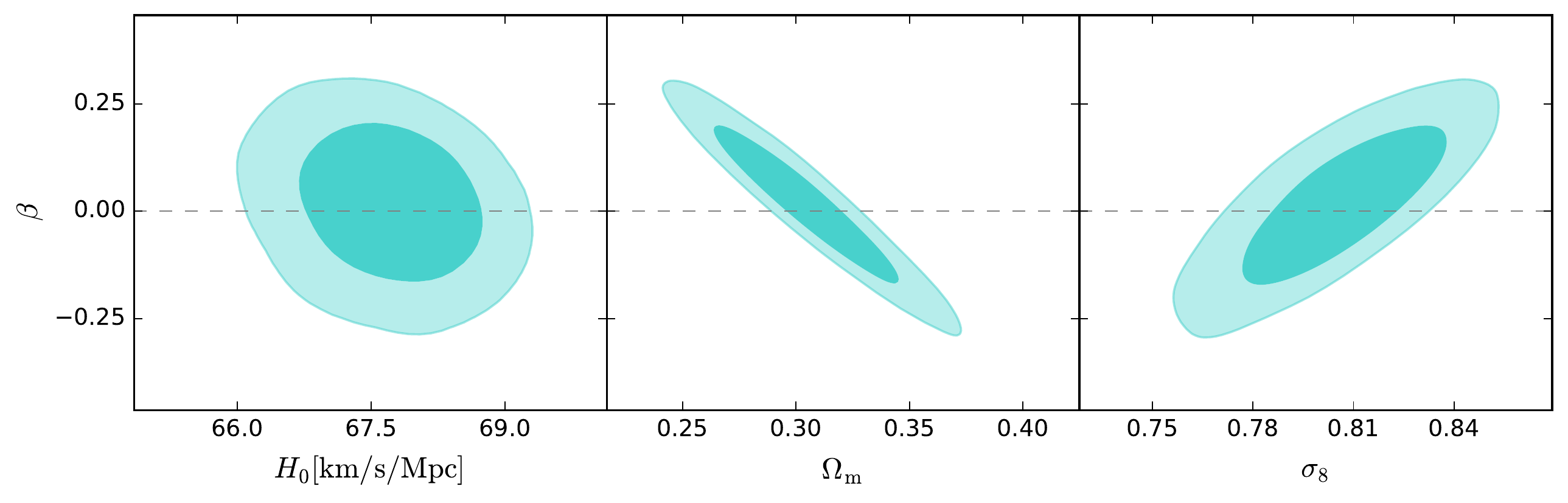}
\caption{The 68\% and 95\% C.L. contours in the planes of $w$ versus $H_{0}$, $\Omega_{\rm m}$ and $\sigma_{8}$ for the $w$CDM model {(green region)}, $\log_{10}B_{0}$ versus $H_{0}$, $\Omega_{\rm m}$ and $\sigma_{8}$ for the $f(R)$ model {(blue region)}, $\beta$ versus $H_{0}$, $\Omega_{\rm m}$ and $\sigma_{8}$ for the I$\Lambda$CDM1 ($Q=\beta H \rho_{\rm c}$) model (purple region), and $\beta$ versus $H_{0}$, $\Omega_{\rm m}$ and $\sigma_{8}$ for the I$\Lambda$CDM2 ($Q=\beta H\rho_{\Lambda}$) model (cyan region).}
\label{fig:all-model}
\end{center}
\end{figure*}

Finally, we discuss how the model parameters affect other cosmological parameters. In Fig.~\ref{fig:all-model}, we show the $68\%$ C.L. and $95\%$ C.L. contours for the specific parameters in different models with late-time parameters $H_{0}$, $\Omega_{\rm m}$, and $\sigma_{8}$ respectively. From this figure, we can see that the parameter $w$ in $w$CDM model can slightly affect $H_{0}$, and the coupling constant $\beta$ in I$\Lambda$CDM2 model can significantly affect $\Omega_{\rm m}$ and $\sigma_{8}$.  However, for the $f(R)$ and \ILCDM1 models, the model parameters of $\log_{10}B_{0}$ and $\beta$ cannot impose evident effects on the other cosmological parameters, because they are tightly constrained by the current observations.


\section{Conclusion}\label{sec:discussion}

In this paper, we discuss the cosmological measurement of the total mass of neutrinos in five typical extensions to the $\Lambda$CDM cosmology including the simplest dynamical dark energy model ($w$CDM), the modified gravity model ($f(R)$), and the interacting vacuum energy models (in which two typical interaction forms of $Q=\beta H\rho_{\rm c}$ and $Q=\beta H\rho_{\Lambda}$ are considered). The $\Lambda$CDM model is considered as a reference model in the analysis. To do cosmological fits, we employ the current observational data sets including the CMB temperature and polarization data, the BAO measurement, the SN data, the $H_{0}$ measurement, the RSD data, the WL data with a prior from the analysis result of DES Y1, and the CMB lensing data.

In our analysis, we obtain the upper limits of the neutrino mass of $\sum m_{\nu}<0.22$ eV for the $\Lambda$CDM model, $\sum m_{\nu}<0.33$ eV for the $w$CDM model, $\sum m_{\nu}<0.20$ eV for the $f(R)$ model, $\sum m_{\nu}<0.37$ eV for the I$\Lambda$CDM1 ($Q=\beta H\rho_{\rm c}$) model, and $\sum m_{\nu}<0.22$ eV for the I$\Lambda$CDM2 ($Q=\beta H\rho_{\Lambda}$) model. Compared to the $\Lambda$CDM model, the $w$CDM model and the I$\Lambda$CDM1 model can yield higher upper limits on the total mass of neutrinos, and the $f(R)$ model and the I$\Lambda$CDM2 model prefer well consistent upper limits of the neutrino mass with the case of $\Lambda$CDM (see Figs.~\ref{fig:mnu} and \ref{fig:mnu-om-sig-h0}).
We also try to discuss how to explain these results (see Fig.~\ref{fig:all-mnu}). We find that the upper limits of the neutrino mass are largely affected by the model parameters. When the model parameters  {(such as $w$ in the $w$CDM model and $\beta$ in the I$\Lambda$CDM1 model)} are in strong correlations with the neutrino mass, the limits of the neutrino mass would have a clear shift, compared to the $\Lambda$CDM model. However, when the model parameters {(such as $B_{0}$ in the $f(R)$ model and $\beta$ in the I$\Lambda$CDM2 model)} are tightly constrained by the current observations, leading to the models approaching to the $\Lambda$CDM model, the obtained constraint limits of the neutrino mass will be well consistent with that in the case of $\Lambda$CDM.


Finally, we study in detail the constraint results of the specific model parameters. {We obtain $w=-1.042^{+0.071}_{-0.024}$ ($1\sigma$) for the $w$CDM model, $\log_{10}B_{0}<-5.67$ ($2\sigma$) for the $f(R)$ model, $\beta=0.0018\pm0.0002$ ($1\sigma$) for the I$\Lambda$CDM1 model, and $\beta=0.0175^{+0.1231}_{-0.1133}$ ($1\sigma$) for the I$\Lambda$CDM2 model. These results indicate that the deviations from the $\Lambda$CDM model are at about $1\sigma$ level, which implies that the current cosmological observations do not favor a deviation from the $\Lambda$CDM cosmology.}

\acknowledgments

This work was supported by the National Natural Science Foundation of China under Grants Nos.~11975072, 11875102, 11835009, and 11690021, and the National Program for Support of Top-Notch Young Professionals.

{}

\end{document}